\documentclass{aa}  
\usepackage{graphicx}
\usepackage{rotating}
\usepackage{amsmath}
\usepackage{txfonts}
\usepackage{natbib}
\bibpunct{(}{)}{;}{a}{}{,}
\begin{document}

\authorrunning{A. Galli \& L. Piro}
\titlerunning{GRB very high energy emission in the context of the ES model}

   \title{Prospects for detection of very high-energy emission from GRB in the context of the external shock model
}

\author{A. Galli\inst{1}
	  \and L. Piro\inst{1}
          }

   \offprints{A. Galli: alessandra.galli@iasf-roma.inaf.it}

   \institute{IASF-Roma/INAF, via fosso del cavaliere 100, 00133 Roma, Italy}

   \date{Received ---; accepted ---}

\abstract {The detection of the 100 GeV-TeV emission by a gamma-ray burst (GRB) will provide an unprecedented opportunity to study the nature of the central engine and the interaction between the relativistic flow and the environment of the burst's progenitor. In this paper we show that there are exciting prospects of detecting from the burst by MAGIC high-energy (HE) emission during the early X-ray flaring activity and, later, during the normal afterglow phase. We also identify the best observational strategy, trigger conditions and time period of observation. We determine the expected HE emission from the flaring and afterglow phases of GRBs in the context of the external shock scenario and compare them with the MAGIC threshold. We find that an X-ray flare with the average properties of the class can be detected in the 100 GeV range by MAGIC, provided that $z \lesssim$0.7. The requested observational window with MAGIC should then start from 10-20 s after the burst and cover about 1000-2000 s. Furthermore, we demonstrate that there are solid prospects of detecting the late afterglow emission in the same energy range for most of the bursts with $z\lesssim$0.5 if the density of the external medium is $n \gtrsim$ a few cm$^{-3}$. In this case, the MAGIC observation shall extend to about 10 - 20 ks. We provide recipes for tailoring this prediction to the observational properties of each burst, in particular the fluence in the prompt emission and the redshift, thus allowing an almost real time decision procedure to decide whether to continue the follow-up observation of a burst at late times.
   
}
   \maketitle
%

\section{Introduction}

Prompt and afterglow emissions in gamma-ray bursts (GRBs) show different temporal and spectral properties and are usually attributed to different mechanisms, more specifically, internal and external shocks, respectively. The internal shocks (IS) occur within the relativistic outflow released after the burst explosion, and an external shock (ES) starts to develop when the ejecta expands in the external medium. The shock with the external medium develops through two components, a forward shock (FS) propagating in the external medium ahead of the expanding shell, and a reverse shock (RS) moving back into the shell itself \citep{meszaros93}. During its expansion into the external medium the fireball collects an increasing amount of the external material. When the inertia of this material equals the fireball energy, the fireball starts to decelerate. The properties of the FS and the RS strongly depend on the GRB duration T$_{GRB}$ with respect to the fireball deceleration time t$_{dec}$. In the standard scenario (i.e. thin shell fireball) the burst duration is shorter than the fireball deceleration time. In this case the RS ends crossing the outflow before the fireball starts to decelerate, thus the fireball releases most of its energy to the external medium, and the ES reaches the peak of its emission at t$_{dec}$. This ES produces synchrotron radiation - i.e afterglow emission - in the X-ray-to-MeV range. This radiation then can be up-scattered by inverse Compton (IC) into the GeV-TeV range \citep{esin01,zhang01}, i.e in the observational band of MAGIC. Very interestingly it has been shown that IC from the afterglow of a ''standard '' thin shell fireball could explain the delayed ($\sim$ 5000 sec after the burst) HE (MeV-to-GeV) emission observed by EGRET in \object{GRB 940217} \citep{galli07}. In this paper we show that MAGIC should be able to the detect the late HE emission from afterglow of GRBs.

The prompt-to-afterglow transition phase in GRBs is characterised by a variety of temporal and spectral behaviours due to the contribution of both prompt and afterglow emissions. In particular this phase, which goes from hundreds-to-thousands of sec after the burst, is characterised by the presence of X-ray flares. Flares are a very common phenomenon being detected in the 30-40\% of the Swift GRB sample (e.g. \citet{falcone07}), and their presence has deep implications because it would imply that the central engine activity is not impulsive as initially thought, but extends on long time scales. Despite the large number of flare models that have been proposed in the literature, none of them has completely interpreted the flare phenomenology. In particular, in the framework of models requiring a long duration central engine activity, X-ray flares could be produced by late internal shocks (LIS) (e.g. \citet{burrows05}; \citet{wu06}; \citet{galli08}) or by a delayed external shock (DES) \citep{piro05,galli06}. In this paper we focus our attention on the possibility that X-ray flares are originated by a DES. 

A DES occurs when the burst duration T$_{GRB}$ is larger than the fireball deceleration time t$_{dec}$ (this corresponds to thick shell fireballs). In this case the RS crosses the shell around T$_{GRB}$ and has the time to become relativistic. As a consequence only a small fraction of the fireball kinetic energy is released to the external medium around t$_{dec}$, and the most of the energy is converted around T$_{GRB}$, i.e. the afterglow emission peaks at T$_{GRB}$. In this context the flare is thus produced by an ES caused by an energy injection lasting until the time of the flare occurrence t$_{f}$, i.e. t$_{f} \sim$ T$_{GRB}$. The DES can straightforwardly explain those X-ray flares which spectrum does not evolve with time and is consistent with that of the afterglow emission. It can also account for flares presenting a spectral evolution of the order of 0.5-1.0 (this value depends on the spectral index of the electron population), if the characteristic emission frequency is crossing the observational energy band. However, in some cases, the spectra of X-ray flares present an hard-to-soft evolution similar to that observed during the prompt emission. This kind of flares could be better explained by LIS. In addition several bursts are characterised by the presence of multiple flares in their X-ray light curve, e.g. \object{GRB 060714} \citep{krimm07}. In such a case a DES could explain only one flare, that representing the onset of the afterglow emission, and the other flares have to be attributed to other mechanisms, for example LIS.

As in the case of a standard ES, also X-ray flare photons can be IC up-scattered giving rise to an HE counterpart peaking in the MeV-to-TeV band, which could be potentially detected by MAGIC. Simultaneous observations of X-ray and HE flares play an important role in order to validate and discriminate flare models, and thus to shed light on the central engine activity.

In this paper we explore the set of conditions that maximise IC emission because we want identify the optimal candidates for a HE emission detection by MAGIC. First, for what regards specifically the flares, we assume the case of DES, which it is generally more favourable to HE emission in comparison with LIS. This is because in the first scenario the peak of IC emission is expected to be at higher energies with respect to the LIS \citep{wang06}. In addition in the DES the emission region is at larger radii in comparison with the LIS, and as a consequence HE emission is less depleted by internal absorption due to pair production. However, it is right to note that in the framework of the LIS model, the External Inverse Compton component (the FS electrons up-scatter on the flare photons) may also peak in GeV band \citep{fan08}. One can distinguish between the DES and the EIC cases looking at temporal behaviour of the low and HE flares. In fact, in the DES scenario one expects a good temporal correlation for the X-ray and HE flares, while in the EIC scenario the HE flare is delayed and much longer in comparison with the X-ray flare (for details see \citet{wang06}).

Then, both for the flares and the late afterglow, we consider the case in which the fireball is in fast cooling regime. In the case of X-ray flare this is well verified. For the late afterglow we derive the conditions on the parameters, particularly on the density $n$ of the external medium, that satisfy this assumption. For the range of parameters explored in this paper the IC emission occurs in the Thompson regime (see for example \citet{wang06} and \citet{galli07}). 

We present the prospects for MAGIC of detecting HE emission during the flare and afterglow phases of GRB in Sect. \ref{flares} and Sect. \ref{afterglow} respectively, and summarise our findings in Sect. \ref{conclusions}.

\section{Trigger conditions and observing time window for high energy emission}
\label{}

\subsection{X-ray flares}
\label{flares}

In the DES scenario the X-ray flare emission can be estimated by analytical equations depending on the following parameters: the fireball kinetic energy $E$, the fractions of energy going respectively into relativistic electrons $\epsilon_e$ and magnetic field $\epsilon_B$, the density of the external medium $n$, the spectral index of the energy distribution of electrons population $p$, and the burst redshift $z$. We assume that at the time of the flare appearance the fireball is in fast cooling (FC) regime. In fact, it can be demonstrated from Eq. \ref{tempo_fs2} of Sect. \ref{afterglow} that for average values of X-ray flare properties the transition from FC to slow cooling (SC) regime occurs at several thousands of sec after the burst. We also assume that the peak of the synchrotron emission $\nu_p$ (i.e. the peak of the X-ray flare in the $\nu$ F$_{\nu}$ space) is below 1 keV (this is in good agreement with the spectral analysis of Swift X-ray flares performed by \citet{falcone07}). Under these two conditions the X-ray flare emission and its peak energy do not depend on the density $n$ of the external medium (see for example \citet{panaitescu00}). The fireball kinetic energy $E$ can be derived from a broad-band spectral fitting. However the aim of this paper is to provide a quick assessment of detectability of HE flares by MAGIC. We thus make the "standard" assumption that the fireball radiative efficiency is of the order of 10\%, i.e. $E \sim 0.1 E_{iso}$. The energy index $p$ of the electrons population can be inferred from the X-ray spectrum, and the burst redshift can be determined through optical spectroscopy. Thus only two parameters remain to be constrained, nominally $\epsilon_e$ and $\epsilon_B$. Finally, these two parameters can be estimated by means of two observational quantities, i.e. the X-ray flare flux density F$_\nu$ and its peak energy $\nu_p$.
It is then possible to link the properties of the HE flare counterpart to observational quantities. By adopting the prescriptions of \citet{panaitescu00} we find that for typical values of the X-ray flare properties the IC parameter is $Y >$ 1. Therefore, the peak energy $\nu_{p,IC}$ and flux $\nu_{IC}$F$_{\nu,IC}$ of the HE flare at the observational frequency $\nu_{obs}$ are expressed by the following relations: 

\begin{equation}
\begin{split}
\nu_{p,IC}& \sim 70 D_{l,28}^{-8/3} (1+z)^{17/12} n^{-1/4} E_{53}^{11/12} \bigg( \frac{\nu_p}{100 eV} \bigg)^{8/3} \bigg( \frac{F_{1keV}}{0.25 mJy} \bigg)^{-4/3}\\
     & \quad \bigg( \frac{t_{obs}}{500s} \bigg)^{-1/12} ~ GeV             
\end{split}
\label{picco_flare}
\end{equation}

\begin{equation}
\begin{split}
\nu_{IC} F_{\nu,IC}& \sim 1.2 \times 10^{-8} e^{-\tau} D_{l,28}^{-4} (1+z)^{-47/48} n^{-1/16} E_{53}^{25/16} \bigg( \frac{\nu_p}{100 eV} \bigg)^{3/2}\\
  & \quad \bigg( \frac{F_{1keV}}{0.25 mJy} \bigg)^{-1} \bigg( \frac{v_{obs}}{100 GeV} \bigg)^{-1/4} \bigg( \frac{t_{obs}}{500 s} \bigg)^{-11/16} ~ erg cm^{-2} s^{-1}
\end{split}
\label{flusso_flare}
\end{equation}

for $\nu_{obs} > \nu_{p,IC}$, and 

\begin{equation}
\begin{split}
\nu_{IC} F_{\nu,IC}& \sim 2 \times 10^{-8} e^{-\tau} D_{l,28}^{-2} (1+z)^{-49/24} n^{1/8} E_{53}^{7/8} \\ 
& \quad \bigg( \frac{\nu_p}{100 eV} \bigg)^{-1/2} \bigg( \frac{v_{obs}}{100 GeV} \bigg)^{1/2} \bigg( \frac{t_{obs}}{500 s} \bigg)^{-5/8}~ erg cm^{-2} s^{-1}
\end{split}
\label{flusso_flare_sottopicco}
\end{equation}

for $\nu_{obs} < \nu_{p,IC}$. In these relations all the quantities are expressed in unity of average values of the class (as an example the flare peak energy typically varies between 10 eV and 1 keV), and we have assumed an electron population spectral index of $p$=2.5. We have also for IC emission the same spectral shape of the synchrotron emission.
The factor e$^{-\tau}$ in Eq.s \ref{flusso_flare} and \ref{flusso_flare_sottopicco} is the attenuation due to the absorption of the extragalactic background (external absorption), with the optical depth $\tau$ varying with the energy of the emitted photons and the redshift of the source. In this paper we evaluate $\tau$ through the analytical model derived by \citet{stecker06}. In particular we find $\tau$(E=100GeV,z) $\sim$0.45 at $z$=0.3, and $\tau$(E=100GeV,z) $\sim$ 8.2 at $z$=1 which implies a cutoff energy of $\sim$ 25 GeV at $z$=1. In the DES model the cutoff energy due to internal pair production (internal absorption) is of the order of 500 GeV for typical model parameters and a burst at $z$=1, and moves at higher energies for nearest events (see for example \citet{galli07}). The attenuation of the high energy ($\sim$ 100 GeV) flux thus it is due mostly to the external absorption.
The time window interesting for the detection of HE emission from flares goes from about 100 s to 2000-3000 s after the burst, that is the time window currently covered by MAGIC observations. We estimate the MAGIC threshold from the upper limits published in \citet{albert07}. The tightest MAGIC upper limit around 100 GeV is of the order of $6 \times 10^{-11}$ erg cm$^{-2}$ s$^{-1}$ for an integration time of 1800 sec. At this time the telescope is background dominated, and we thus estimate the MAGIC threshold at 100 GeV to be of the order of 1.1 $\times$10$^{-10}$ erg cm$^{-2}$ s$^{-1}$ for an integration time of 500 s. The upper limits reported in \citet{albert07} are at 95\% confidence level, and this implies that the MAGIC threshold we have estimated is at roughly 1 $\sigma$. Wecderive from Eq. \ref{flusso_flare} and \ref{flusso_flare_sottopicco} that MAGIC would be able to detect HE emission from flares up to $z \sim$ 0.7 if the X-ray flux and peak energy are at the mean values.

Finally, one can note from Eq.s \ref{picco_flare}, \ref{flusso_flare} and \ref{flusso_flare_sottopicco} that the peak energy and flux of the HE flare depend weakly on the density $n$. Consequently, these estimates are not strongly affected by our uncertainties on the density of the external medium. It is also important to stress that Eq. \ref{flusso_flare} shows that the flux of the HE flare increases with lower values of the flux density of the X-ray flare. This is because during the FC regime and above the peak of the emission, all the energy stored into relativistic electrons is radiated away, thus if the energy emitted through the synchrotron channel decreases the energy available to the IC channel has to increase.

\subsection{The standard external shock model}
\label{afterglow}

\begin{figure}
\centering
\includegraphics[width=7cm,angle=-90.0]{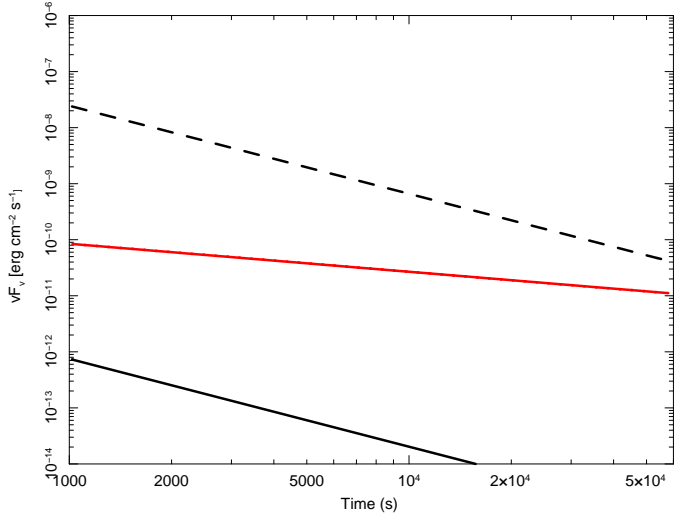}
\caption{IC afterglow light curves at 100 GeV for a burst with typical energy and afterglow luminosity at $z$=1.0 (black solid line) and $z$=0.3 (black dashed). The HE afterglow light curves are corrected for the attenuation due to the absorption of the extragalactic background. The red line is the MAGIC threshold at 100 GeV as a function of the integration time.}
\label{fig_lumconst_afterglow}
\end{figure}

Afterglow emission from a “standard” ES is satisfactorily described by analytical equations depending on the parameters $E$, $\epsilon_e$, $\epsilon_B$, $n$, $p$, and $z$, as in the case of DES. We assume that the afterglow emission occurs in FC regime because it can be shown that for a large range of values of the ES parameters, the transition from the FC to the SC regime can take place at very late times (we will show later the range of applicability of this assumption). Then we find that, for the range of parameters explored, $Y$ is always $>$ 1. It can be demonstrated that, adopting the same procedure applied to X-ray flares, the peak energy and flux of the IC counterpart of the X-ray afterglow are expected to be:

\begin{equation}
\begin{split}
v_{p,IC}& \sim 2.5 D_{l,28}^{-8/3}(1+z)^{17/12} n^{-1/4} E_{53}^{11/12} \bigg( \frac{\nu_p}{2eV} \bigg)^{8/3} \bigg( \frac{F_{1keV}}{1\mu Jy} \bigg)^{-4/3}\\
   & \quad \bigg( \frac{t_{obs}}{10^4s} \bigg) ^{-1/12} ~ GeV                         
\end{split}
\label{picco_afterglow}
\end{equation}

\begin{equation}
\begin{split}
\nu_{IC} F_{\nu,IC}& \sim 10^{-9} e^{-\tau} D_{l,28}^{-4}(1+z)^{-47/48} E_{53}^{25/16} n^{-1/16} \bigg( \frac{\nu_p}{2eV} \bigg)^{3/2}\\
    & \quad \bigg( \frac{F_{1keV}}{1 \mu Jy} \bigg)^{-1} \bigg( \frac{\nu_{obs}}{100 GeV} \bigg)^{-1/4} \bigg( \frac{t_{obs}}{10^4s} \bigg)^{-11/16} ~ erg cm^{-2}s^{-1}               
\end{split}
\label{flusso_afterglow}
\end{equation}

for $\nu_{obs} > \nu_{p,IC}$. Note that the HE afterglow flux is corrected for the external absorption, and now we have normalised the observable properties of the afterglow to their average values around 10 ks after the burst. By following the same approach adopted in the preceding section we estimate the 100 GeV MAGIC threshold to be $\sim$ 3$\times$10$^{-11}$ erg cm$^{-2}$ s$^{-1}$ for an integration time of 10 ks. 
In Fig. \ref{fig_lumconst_afterglow} we compare the late HE emission from the afterglow of a GRB exploded at redshift $z$=1 (black solid line) and from a low distance burst at $z$=0.3 (black dashed line) with the 100 GeV MAGIC threshold (red solid line). We assume that the intrinsic properties of the burst are the same at $z$=1 and $z$=0.3.
The HE afterglow flux produced by a burst at $z$=1 is heavily depleted by the external absorption, that makes it undetectable by MAGIC. On the contrary the HE flux expected from a burst located at $z$=0.3 is weakly affected by the external absorption and, as shown in Fig. \ref{fig_lumconst_afterglow}, it can be detected even at times larger than 10 ks.

We note that the expected flux at late times can be derived by extrapolating the early afterglow behaviour, and that one can use this information to get a trigger criterion based on early-time observations (around 1000 sec) of each burst. From Fig. \ref{fig_lumconst_afterglow} it is clear that a good candidate for late observations will also show a prominent HE flux at early time, so MAGIC itself can provide the information needed to decide to which time extend the observation. Alternatively one should use the early time X-ray and optical information to predict the HE flux. To this aim is important to check that the X-ray afterglow emission is dominated by the ES, i.e. that the afterglow light curve has attained the typical power law decay. 

\begin{figure}
\centering
\includegraphics[width=7.0cm,angle=90.0]{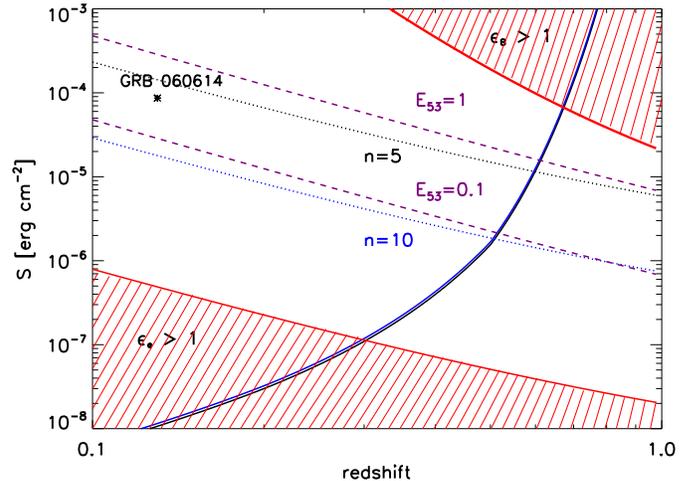}
\caption{Requirements on the burst fluence $S$ and on the external medium density $n$ for the detectability of HE emission (100 GeV) from the afterglow of a GRB around 10 ks, as a function of the redshift $z$. We assume the peak frequency of the X-ray afterglow to be $\nu_p$=2 eV, and that the afterglow luminosity is proportional to the energy $E_{iso}$. Solid and dotted lines represent the fluence lower limits deriving from Eq. \ref{trigger_limit} and Eq. \ref{fastcooling_limit} respectively. We present these lower limits for $n$=5 (black) and $n$=10 (blue). This figure shows that a bright and low distance burst as \object{GRB 060614} could be detected by MAGIC if $n \gtrsim$ 6. As a reference we also plot as a function of the redshift the fluence corresponding to a burst with $E_{53}$=1 and $E_{53}$=0.1.}
\label{fig_detection}
\end{figure}

Here we provide a more general and easier-to-apply trigger condition derivable immediately from the fluence of the prompt emission and on the redshift, that holds for reasonable early afterglow properties. In deriving this trigger criterion we assume that the X-ray afterglow luminosity is proportional to the burst isotropic energy $E_{iso}$. This it has been observed in the BeppoSAX GRB X-ray afterglow catalogue \citep{depasq06}, and it is expected theoretically \citep{freedman01}. 
We scale the X-ray afterglow luminosity to gamma-ray isotropic energy ratio observed in the GRB BeppoSAX catalogue by \citet{depasq06} at 40 ks after the burst (rest frame), at 10 ks taking into account that above the peak of the emission the afterglow luminosity goes as $t^{-(3p-2)/4}$. Thus with a spectral index of the electron population $p=2.5$ we find that around 10 ks after the burst $L_x \sim 7.53 \times 10^{-7} E_{iso}$, where the X-ray afterglow luminosity is at 1 keV and the burst isotropic energy $E_{iso}$ is in the 40-700 keV energy range. We finally assume a radiative efficiency of 10\% to convert the fireball kinetic energy $E$ into the burst isotropic energy $E_{iso}$. Under these assumptions Eq. \ref{flusso_afterglow} becomes:

\begin{equation}
\begin{split}
\nu_{IC} F_{\nu,IC}& \sim 3 \times 10^{-7} e^{-\tau} D_{l,28}^{-7/8}(1+z)^{-37/24} S^{9/16} n^{-1/16} \\
    & \quad \bigg( \frac{\nu_p}{2eV} \bigg)^{3/2} \bigg( \frac{\nu_{obs}}{100 GeV} \bigg)^{-1/4} \bigg( \frac{t_{obs}}{10^4s} \bigg)^{-11/16} ~ erg cm^{-2}s^{-1}               
\end{split}
\label{flusso_afterglow_fluence}
\end{equation}

where $S$ is the burst bolometric fluence (we remember that $S$=(1+z)E$_{iso}$/4 $\pi D_l^2$). By comparing Eq. \ref{flusso_afterglow_fluence} with the threshold of MAGIC at 100 GeV for an integration time of 10 ks we derive the following lower limit on the burst fluence $S$:

\begin{equation}
\begin{split}
S& \gtrsim 7.7 \times 10^{-8} \big( e^{-\tau} \big)^{-16/9} D_{l,28}^{14/9}(1+z)^{74/27} n^{1/9} \bigg( \frac{\nu_p}{2eV} \bigg)^{-8/3} \\
& \quad \bigg( \frac{\nu_{obs}}{100 GeV} \bigg)^{4/9} \bigg( \frac{t_{obs}}{10^4s} \bigg)^{11/9} ~ erg cm^{-2}
\end{split}
\label{trigger_limit}
\end{equation}

where we have assumed the average values of the observational properties of the afterglow emission. We note that Eq. \ref{trigger_limit} depends weakly on the density $n$ of the external medium, and plot in Fig. \ref{fig_detection} this trigger requirement for some values of $n$ in the range of redshift of interest for MAGIC (0.1 $ \lesssim z <$ 1.0). It is important to note that the trigger requirement given by Eq. \ref{trigger_limit} holds only if at the observation time $t_{obs}$ the fireball is in FC regime. Following the prescriptions of \citet{panaitescu00} the fast-to-slow cooling transition time $t_{FS}$ as a function of the of the parameters of the "standard" ES model is:

\begin{equation}
t_{FS} \sim 2150 (1+z) E_{53} n (1+Y)^2 \epsilon_{e,-1}^2 \epsilon_{B,-1}^2 ~ s
\label{tempo_fs}
\end{equation}

We follow the same approach adopted to derive Eq.s \ref{picco_flare}, \ref{flusso_flare}, \ref{picco_afterglow} and \ref{flusso_afterglow}, and write the transition time $t_{FS}$ as a function of the observational properties of the afterglow emission:

\begin{equation}
\begin{split}
t_{FS}& \sim 2619 (1+z)^{-1/3} E_{53}^{-1/3} n \bigg( \frac{\nu_p}{2 eV} \bigg)^{7/6} \bigg( \frac{L_{1keV}}{7.25 \times 10^{45} erg s^{-1}} \bigg)^{2/3}\\
& \quad \bigg( \frac{t_{obs}}{10^4 s} \bigg)^{8/3} ~ s.
\end{split}
\label{tempo_fs2}
\end{equation}

Finally we write the afterglow luminosity as proportional to the burst energy, and link the energy to the burst fluence $S$:

\begin{equation}
t_{FS} \sim 1.3 \times 10^{5} (1+z)^{-2/3} D_{l,28}^{2/3} n S^{1/3} \bigg( \frac{\nu_p}{2 eV} \bigg)^{7/6} \bigg( \frac{t_{obs}}{10^4 s} \bigg)^{8/3} ~ s
\label{tempo_fs3}
\end{equation}

Then we require that the transition from FC to SC regime occurs at times larger then 10 ks thus obtaining a second lower limit on the burst fluence $S$:

\begin{equation}
S \gtrsim 4.4 \times 10^{-4} D_{l,28}^{-2}(1+z)^{2} n^{-3} \bigg( \frac{\nu_p}{2eV} \bigg)^{-7/2} \bigg( \frac{t_{obs}}{10^4s} \bigg)^{-8}~ erg cm^{-2}
\label{fastcooling_limit}
\end{equation}

where we have normalised again the properties of the afterglow emission to their average values. This limit strongly depends on the density of the external medium $n$, in particular high densities favour the FC regime. In conclusion if an event satisfies both the fluence requirement given by Eq. \ref{trigger_limit} and that given by Eq. \ref{fastcooling_limit}, it can produce HE emission detectable by MAGIC.
We summarise the condition of detectability and applicability of the trigger criterion in Fig. \ref{fig_detection}. In this figure we plot the fluence lower limits deriving from the 100 GeV MAGIC threshold (solid lines) and from the request that at the observation time t$_{obs}$ the fireball is in FC regime (dotted lines), as a function of the burst redshift for different values of the density of the external medium $n$. Note that the limits on the fluence deriving from the MAGIC threshold increases quickly at redshift $z \sim$ 0.5 because of the extragalactic absorption.We also indicate in Fig. \ref{fig_detection} the forbidden regions of fluences and redshift implying values of $\epsilon_e$ and $\epsilon_B$ larger than 1 (red dashed regions).

If the density of the external medium is $n \gtrsim$ 200 the fireball is always in FC regime, thus all the events above the solid lines of Fig. \ref{fig_detection} can be detected by MAGIC, and the region of fluences and redshifts below these lines is forbidden for MAGIC. For lower values of $n$ the range of redshift values where the FC condition is satisfied decreases, and as a consequence also the fluence lower limits associated to the MAGIC threshold can be applied only in this restricted range of redshifts. In particular we cannot say anything about the possibility for MAGIC to detect HE emission from those event that are above the fluence limit related to the MAGIC threshold but below the FC limit because in this case the fireball is in SC regime (we will present the case when the fireball is in SC regime in a future paper). As a reference a burst with $E_{53}$=1 can be detect by MAGIC up to $z \sim$ 0.6 if $n \gtrsim$5, and a burst with $E_{53}$=0.1 can be detect by MAGIC up to $z \sim$ 0.5 if $n \gtrsim$10. A bright event such as the low distance \object{GRB 060614} at $z$=0.13 (redshift estimated from emission lines due to the GRB host galaxy; \citet{price06}), having $S(20 keV-2 MeV) \sim 4 \times 10^{-5}$ erg cm$^{-2}$ and $S(1 keV-10 MeV) \sim 8.6 \times 10^{-5}$ erg cm$^{-2}$ \citep{golenetskii06}, could be detected by MAGIC if $n \gtrsim 6 cm^{-3}$ \footnote{Note that the reference to \object{GRB 060614} would be just an example aimed to show that also a low redshift burst can have a large fluence and thus can be potentially detect by MAGIC up to late times. In fact, we note that the afterglow emission of \object{GRB 060614} settles on the ''standard'' ES model after 30-40 ksec \citep{mangano07}.}.

Finally we estimate how much the predicted HE emission is above the MAGIC threshold if both the requirements expressed by Eq.s \ref{trigger_limit} and \ref{fastcooling_limit} are satisfied. Equation \ref{flusso_afterglow_fluence} shows that the IC afterglow flux goes approximately as the square root of the burst fluence, and this implies that if, for example $S$ is a factor 10 above the trigger requirement \ref{trigger_limit} then the IC flux is expected to be about a factor 3 above the MAGIC threshold. 

\section{Discussion and conclusions}
\label{conclusions}

In this paper we study the possibilities for MAGIC to detect HE emission during the flare and afterglow phases of GRBs. We adopt the set of conditions that maximise IC emission in order to include all the potential events that can be detected by MAGIC. 
Concerning X-ray flares the optimal conditions is the DES scenario, because it predicts larger IC fluxes and higher peak energies in comparison with the LIS scenario. We find that by considering the average properties (peak energy and flux) of the population of X-ray flares, MAGIC would be able to detect HE flares up to $z \sim$ 0.7. We thus suggest that the best observational strategy to detect with MAGIC HE emission from flares is to observe all bursts with $z \lesssim$ 0.7 for 1000-2000 sec, the temporal range of flares occurrence.

For the afterglow phase of GRB the optimal condition is FC regime, even at late times as 10-20 ks. For average values of the observed afterglow properties this applies if $n \gtrsim$ a few cm$^{-3}$. Then we expect that, by assuming the average properties of the class, bursts at $z \lesssim$ 0.5 will likely produce a HE afterglow detectable by MAGIC as late as $\gtrsim$ 10 ks. This would imply a signicative extension of the duration of MAGIC GRB observations, that usually cover the first 1000-2000 sec after the burst. We have thus derived a trigger criterion based on the fluence of the prompt emission, to individuate those GRBs that could have a HE afterglow detectable by MAGIC. In deriving this criterion we assume that the afterglow luminosity is proportional to the burst energy, and derive the minimum burst fluence necessary for the IC flux to be above the MAGIC threshold. In addition, the request that at 10 ks the burst is still in FC regime introduces a second lower limit on the burst fluence. Consequently, only those bursts which satisfy both the conditions can be detected by MAGIC. We find that bursts with fluence $\gtrsim 10^{-7}-10^{-6}$ erg cm$^{-2}$ and redshift $z \lesssim$ 0.5 have good possibilities to produce HE emission detectable by MAGIC if $n$ is $\gtrsim$ few cm$^{-3}$. It is important to note that our predictions hold in the Thompson limit. However, if the energy of the electrons and of the scattering photons is very high (i.e larger than 100 GeV), then the Klein-Nishina process could affect significantly the HE emission. In the Klein-Nishina regime the spectrum of IC component is softer than that predicted in the Thompson regime (for a detailed discussion see \citet{guetta03}). In this case the HE flux it is lower than the one expected in the Thompson regime and thus it will be more difficult to detect (for a comparison of the expected flux with the GLAST threshold see \citet{fan08}).

Our results are consistent with the upper limits derived so far by MAGIC \citep{albert07}. In the sample of 9 events, the 4 events with know distance are at redshift $z >$3. In addition, from the updated Swift GRB redshift distribution (see the web page //www.astro.ku.dk/~pallja/GRBsample.html) we derive that $\sim$ 5\% of Swift GRB with know distance are at redshift $\lesssim$ 0.5, and $\sim$ 10 \% are at $z \lesssim$ 0.7. Therefore we expect that less than 1 event of the MAGIC sample is at $z <$ 0.5 with a probabilty of $\sim$ 63 \%, and that less than 1 event is at $z <$ 0.7 with a probability of $\sim$ 40 \%. We also note that the observations reported in \citet{albert07} end around 2000-3000 s after the burst. In order to test the so called standard external shock afterglow model is thus important that in the future MAGIC observations will extend up to $\sim$ 10-20 ks after the burst, at least for low distance GRBs.




\begin{thebibliography}{} 
\bibitem[Albert et al. (2007)]{albert07} Albert, J., Aliu, E., Anderhub, H., et al., 2007, ApJ, 667, 358
\bibitem[Burrows et al. (2005)]{burrows05} Burrows, D.N., Romano, P., Falcone, A., et al., 2005, Science, 309, 1833
\bibitem[De Pasquale et al. (2006)]{depasq06} De Pasquale, M., Piro, L., Gendre, B., et al., 2006, A\&A, 455, 813
\bibitem[Falcone et al. (2007)]{falcone07} Falcone, A.D., Morris, D., Racusin, J., et al, 2007, ApJ, 671, 1921
\bibitem[Fan et al. (2008)]{fan08} Fan, Y.Z., Piran, T., Narayan, R. \& Wei, D., 2008, MNRAS, 384, 1483
\bibitem[Freedman \& Waxman (2001)]{freedman01} Freedman, D.L., \& Waxman, E., 2001, ApJ, 547, 922
\bibitem[Galli \& Piro (2006)]{galli06} Galli, A., \& Piro, L., 2006, A\&A, 455, 413
\bibitem[Galli \& Piro (2007)]{galli07} Galli, A., \& Piro, L., 2007, A\&A, 475, 421
\bibitem[Galli \& Guetta (2008)]{galli08} Galli, A., \& Guetta, D., 2008, A\&A, 480, 5
\bibitem[Golenetskii et al. (2006)]{golenetskii06} Goleneteskii, S., Aptekar, R., Mazets, E., et al., GCN Circ. 5264
\bibitem[Guetta \& Granot (2003)]{guetta03} Guetta, D., \& Granot, J., 2003, MNRAS, 340, 115
\bibitem[Krimm et al. (2007)]{krimm07} Krimm, H.A., Granot, J., Marshall, F.E., et al., 2007, ApJ, 665, 554
\bibitem[Mangano et al. (2007)]{mangano07} Mangano, V., Holland, S.T., Malesani, D., et al., 2007, A\&A, 470, 105
\bibitem[Meszaros \& Rees (1993)]{meszaros93} Meszaros, P., \& Rees, M., 1993, 
\bibitem[Panaitescu \& Kumar (2000)]{panaitescu00} Panaitescu A. \& Kumar P., 2000, ApJ, 543, 66
\bibitem[Piro et al. (2005)]{piro05} Piro, L., De Pasquale, M., Soffitta, P., et al., ApJ, 623, 314
\bibitem[Price et al. (2006)]{price06} Price, P.A., Berger, E., \& Fox, D.B., GCN Circ. 5275
\bibitem[Sari \& Esin (2001)]{esin01} Sari, R., \& Esin, A.A., 2001, ApJ, 548, 747
\bibitem[Stecker et al. (2006)]{stecker06} Stecker, F.W., Malkman, M.A., \& Scully, S.T., 2006, ApJ, 648, 774
\bibitem[Wang et al. (2006)]{wang06} Wang, X.Y., Li, Z.,  \& Meszaros, P., 2006, ApJ, 641, L89
\bibitem[Wu et al. (2006)]{wu06} Wu, X.F., Dai, Z.G., Wang, X.Y., et al., 2006, 36th COSPAR Scientific Assembly, p. 731
\bibitem[Zhang \& Meszaros (2001)]{zhang01} Zhang, B., \& Meszaros, P., 2001, \apj, 559, 110

\end{thebibliography}
\end{document}